\documentclass[fleqn,usenatbib]{mnras}

\usepackage{newtxtext,newtxmath}
\usepackage[T1]{fontenc}
\usepackage{ae,aecompl}

\usepackage{graphicx}	% Including figure files
\usepackage{enumerate}
\usepackage{color}
\usepackage{graphicx}
\usepackage{multicol}
\usepackage{savesym}
\usepackage{color}
\usepackage{tipa}
\usepackage{csvsimple}
\usepackage{arydshln}
\usepackage{arydshln}

\title[System parameters of MAXI J0637$-$430]
{A 2-hr binary period for the black hole transient MAXI J0637$-$430}

\author[R. Soria et al.]
{Roberto Soria$^{1,2}$ \thanks{Email: rsoria@nao.cas.cn (RS)}, 
Ruican Ma$^{3,4}$, Lian Tao$^{3}$, Shuang-Nan Zhang$^{3,4}$
\\
$^{1}$College of Astronomy and Space Sciences, University of the Chinese Academy of Sciences, Beijing 100049, China\\
$^{2}$Sydney Institute for Astronomy, School of Physics A28, The University of Sydney, Sydney, NSW 2006, Australia\\
$^{3}$Key Laboratory of Particle Astrophysics, Institute of High Energy Physics, Chinese Academy of Sciences, Beijing 100049, China\\
$^{4}$University of Chinese Academy of Sciences, Chinese Academy of Sciences, Beijing 100049, People's Republic of China\\
\\
}

\date{Accepted 2022 July 04. Received 2022 June 24; in original form 2022 April 26}

\pubyear{2022}

\begin{document}
\label{firstpage}
\pagerange{\pageref{firstpage}--\pageref{lastpage}}
\maketitle

% Abstract of the paper
\begin{abstract}
We revisit various sets of published results from X-ray and optical studies of the Galactic black hole (BH) candidate MAXI J0637$-$430, which went into outburst in 2019. Combining the previously reported values of peak outburst luminosity, best-fitting radii of inner and outer accretion disk, viewing angle, exponential decay timescale and peak-to-peak separation of the He {\footnotesize II} $\lambda 4686$ disk emission line, we improve the constraints on the system parameters. We estimate a heliocentric distance $d \approx (8.7 \pm 2.3)$ kpc, a projected Galactocentric distance $R \approx (13.2 \pm 1.8)$ kpc and a height $|z| \approx (3.1 \pm 0.8)$ kpc from the Galactic plane. It is the currently known Milky Way BH candidate located farthest from the Galactic Centre.
%(corresponding to a location in the Galactic halo or the tail end of the thick disk). I, 
We infer a BH mass $M_1 \approx (5.1 \pm 1.6) M_{\odot}$, a spin parameter $a_{\ast} \lesssim 0.25$, a donor star mass $M_2 \approx (0.25 \pm 0.07) M_{\odot}$, 
%a binary separation $a \approx (1.1 \pm 0.3) \times 10^{11}$ cm, 
a peak Eddington ratio $\lambda \approx 0.17 \pm 0.11$ 
%(just high enough for the system to reach the high/soft state) 
and a binary period $P_{\rm orb} \approx 2.2^{+0.8}_{-0.6}$ hr. This is the shortest period measured or estimated so far for any Galactic BH X-ray binary. If the donor star is a main-sequence dwarf, such a period corresponds to the evolutionary stage where orbital shrinking is driven by gravitational radiation and the star has regained contact with its Roche lobe (low end of the period gap). %Alternatively, the donor may be a more evolved, stripped star. 
The three Galactic BHs with the shortest period ($\lesssim$3 hr) are also those with the highest vertical distance from the Galactic plane ($\gtrsim$2 kpc). This is probably because binaries with higher binding energies can survive faster natal kicks. 
\end{abstract}


\section{Introduction}
MAXI J0637$-$430 is a transient low-mass X-ray binary, discovered by the {\it Monitor of All-sky X-ray Image} ({\it MAXI}) when it went into outburst on 2019 November 2 \citep{negoro19}. Its outburst evolution over the following weeks was monitored by several teams with a variety of X-ray and multiband facilities: the {\it Neil Gehrels Swift Observatory} ({\it Swift}) X-ray Telescope (XRT) and UltraViolet and Optical Telescope (UVOT) \citep{kennea19}, the {\it Insight-HXMT} X-ray observatory \citep{ma22}, the {\it Nuclear Spectroscopic Telescope Array} ({\it NuSTAR}) \citep{tomsick19}, the {\it Neutron star Interior Composition Explorer} ({\it NICER}) \citep{remillard20}, {\it Astrosat} \citep{thomas19}, the Southern Astrophysical Research (SOAR) telescope \citep{strader19}, the Australia Telescope Compact Array (ATCA) \citep{russell19}. Based on this rich set of multiband data, several studies \citep{tetarenko21,lazar21,jana21,ma22} have presented detailed analyses of the properties of this Galactic black hole (BH) candidate.

MAXI J0637$-$430 stands out from other BH transients for several unusual properties: i) it is located at high Galactic latitude, well outside the disk plane, and large Galactocentric distance. If its proposed heliocentric distance estimate ($d\sim 7$ kpc: \citealt{ma22}) is correct, it may belong to the thick disk population and may have a different formation history than most other Galactic BH transients; ii) also if the estimated distance is correct, the peak bolometric luminosity $L_{\rm bol}$ of the 2019 outburst was only $L_{\rm bol} \sim 0.1 L_{\rm Edd}$, and the system remained in the thermal dominant state during outburst decline at or just below $L_{\rm bol} \sim 0.01 L_{\rm Edd}$; iii) the outer radius of the accretion disk is relatively small, which implies a small binary separation and very short orbital period ($P \sim 2$--4 hr: \citealt{tetarenko21}). In fact, it was suggested \citep{ma22} that MAXI J0637$-$430 has the shortest period known to-date among transient BH candidates \citep{knevitt14,Corral-Santana2016,tetarenko16,arur18}. 

There is no reported detection of the donor star in quiescence: therefore, there is no dynamic mass estimate for this system. The distance estimates so far are also rather speculative. Both uncertainties hamper our efforts to model the physics of MAXI J0637$-$430 and place it in context of binary stellar evolution. For example, if the distance was somewhat larger, MAXI J0637$-$430 would be the first BH X-ray binary detected in the Galactic halo. The unusually low color temperature of the inner accretion disk ($kT_{\rm in} \approx 0.70$ keV) at peak outburst may be due to a low Eddington ratio, but it could also signal a more massive BH located at a higher distance. 

To reduce this uncertainty on BH mass, distance, binary period and type of donor star, we have collected and re-examined the main observational evidence and modelling results presented so far in the literature  (Table 1). We will show that, even in the absence of phase-resolved optical spectroscopy, by combining those constraints, we can obtain more precise values for those quantities. The main ingredients for this exercise will be: i) the inner disk radius derived from disk-blackbody model fits to the soft (thermal) X-ray emission. In particularly, the X-ray spectra from {\it Swift}/XRT provide the tightest constraint; ii) the outer disk radius from the broad-band spectral energy distribution provided by the combined dataset of {\it Swift}/XRT and {\it Swift}/UVOT; iii) an independent constraint on outer disk radius and BH mass provided by the double-peaked He {\footnotesize II} $\lambda$4686 emission lines in the Gemini spectra of \cite{tetarenko21}; iv) the bolometric unabsorbed flux at the peak of the outburst. We will combine those empirical parameters with the help of well-tested models of BH accretion: the standard solution for an optically thick, geometrically thin accretion disk \citep{ss73,frank02}; the basic framework of the thermal-viscous instability model \citep{king98,dubus99,lasota01,lasota15,hameury20}, and in particular, its prediction for the peak bolometric luminosity of an outburst; the theoretical profile of disk emission lines \citep{smak81}; the Roche lobe geometry properties as a function of mass ratio \citep{eggleton83}.
%Write later....something about the general properties of this source, and what is still unknown. Explain that there are now four X-ray studies of this BH candidate \citep{tetarenko21,lazar21,jana21,ma22}, based on different instruments and with different choices of spectral models. Those papers have proposed different ranges for the system parameters. Here, we try to estimate those parameters, putting together the information from those studies....

\section{Black hole mass estimate}

\subsection{Mass constraints from the inner-disk emission}
During most of its only recorded outburst, MAXI J0637$-$430 was in the thermal dominant state. Thus, we plausibly assume that, near outburst peak, most of the thermal emission comes from a standard Shakura-Sunyaev disk extended down to the innermost stable circular orbit (ISCO). It was previously noted \citep{lazar21,ma22} that systematic spectral residuals may require a second thermal component; however, the characteristic length scale of the hottest thermal emitter is consistent for various models. We assume that such radius is a good proxy for ISCO. As customary, we distinguish here between an apparent inner-disk radius $r_{\rm in}$, directly derived from the normalization of disk-blackbody models in {\sc xspec} \citep{arnaud96}, and a physical inner disk radius $R_{\rm in} \approx R_{\rm ISCO} \approx \xi f_{\rm col}^2 r_{\rm in}$ \citep{kubota98}, where $\xi \approx 0.412$ is a geometric correction factor, and $f_{\rm col}$ is the hardening factor (colour correction), as defined by \cite{shimura95} (see also: \citealt{zhang97,mcclintock14}). 

A traditional choice of hardening factor is $f_{\rm col} = 1.7$, from \cite{shimura95}: this corresponds to the ``standard'' conversion factor $R_{\rm in} \approx 1.19 r_{\rm in}$.
More detailed theoretical models of disk atmospheres \citep{davis05,davis06,shafee06,done08} suggest a slightly lower value of $f_{\rm col}$ ($1.6 \lesssim f_{\rm col} \lesssim 1.7$) for a range of parameters suitable to MAXI J0637$-$430: intermediate viewing angles, peak colour temperature $\sim$0.7 keV, disk luminosity $L_{\rm disk} \approx 0.1 L_{\rm Edd}$ and low BH spin (as we will show later). Thus, for our BH mass estimate we assume $R_{\rm in} \approx (1.1 \pm 0.1) r_{\rm in}$. The lower boundary of this correction factor (corresponding to $f_{\rm col} \gtrsim 1.55$) is a solid limit, both from theoretical arguments and observational evidence. For the upper boundary, we include the standard value $f_{\rm col} = 1.7$. This limit is not as robust: some studies \citep{gierlinski04,done08} have suggested $f_{\rm col} \sim 1.8$--2.0. Luckily, the uncertainty on the upper limit of $f_{\rm col}$ (corresponding to higher BH masses) does not substantially affect our mass estimates, because it corresponds to a range of BH masses that is ruled out by another empirical constraint (as discussed later, Section 2.2).

The average of the best-fitting values of $r_{\rm in}$ over the first 10 \textit{Swift}/XRT observations after outburst peak (2019 November 9--19)\footnote{We have excluded here the XRT observations on November 6 and November 8 (ObsIDs 12172001 and 12172002, respectively) because of instrumental problems \citep{ma22}.} is $r_{\rm in}\sqrt{\cos \theta} \approx (33.0 \pm 1.3)\, d_{10}$~km when the spectrum is fitted with an absorbed {\tt diskbb} $+$ {\tt powerlaw} model \citep{ma22}; here, $\theta$ is the viewing angle, $d_{10}$ is the distance in units of 10 kpc, and the quoted uncertainty is the 1-$\sigma$ scatter of the 10 best-fitting values. When the same 10 spectra are fitted with a {\tt diskir} model, the average inner radius is $r_{\rm in}\sqrt{\cos \theta} \approx (35.0 \pm 2.0)\, d_{10}$~km \citep{tetarenko21}. In fact, the best-fitting value of $r_{\rm in}$ has an increasing trend through the outburst ({\it e.g.}, Fig.~3 in \citealt{tetarenko21}), which can be interpreted either as the inner truncation radius of the disk gradually moving away from ISCO, or the hardening factor becoming slightly lower as the luminosity decreases (both are, in principle, plausible explanations). Regardless of the reason for the change in the apparent radius, the smallest value of $r_{\rm in}$ measured just after outburst peak is likely to be the best proxy for the true value of $R_{\rm isco}$. This is the value measured on 2019 November 9 (MJD 58796.74, ObsID 00012172003): $r_{\rm in}\sqrt{\cos \theta} \approx (32 \pm 1)\, d_{10}$~km, from both a {\tt diskbb} \citep{ma22} and a {\tt diskir} \citep{tetarenko21} model fit to the XRT data. Finally, the viewing angle is independently known, $\theta = \left(64^{\circ} \pm 6^{\circ}\right)$, from the reflection component fitted to {\it NuSTAR} data \citep{lazar21}.

%An estimate of $i = 64^{\circ} \pm{6}^{\circ}$ was provided by \cite{Lazar2021}, based on the best-fitting parameters of a reflection component in their {\it NuSTAR} spectra (3--79 keV band).

%Our second assumption is that the spectrum can be approximated with a {\it diskbb} model, with the hardening factor and geometric correction factor introduced by \citep{kubota98}. This means that the disk normalization $N_{\rm dbb}$, the viewing angle $\theta$, the distance $d$, fitted (apparent) inner-disk radius $r_{\rm in}$ and the physical inner radius $R_{\rm in}$ are related as $R_{\rm in} \approx 1.19 \, r_{\rm in} \approx 1.19 \left(N_{\rm dbb}/\cos \theta\right)^{1/2}\,(d/10\,{\mathrm {kpc}})$. 
In summary, combining the previous parameters and propagating their uncertainties\footnote{Throughout this work, uncertainty ranges are computed with Gaussian error propagation for uncorrelated variables, except where explictly noted otherwise.}, we obtain a value of $R_{\rm in} \approx R_{\rm ISCO} \approx (51 \pm 6) \, d_{10}$ km.
%$\approx (36 \pm 4) \, d_{7}$ km. Henceforth, for convenience (justified a posteriori), we will normalize distances in units of 7 kpc, following the distance estimate of \cite{ma22}.
%, we will generally normalize the distance $d$ to units of 7 kpc in the rest of our analysis, for convenience ($d_7 \equiv d/7\,{\mathrm {kpc}}$. 
For a Schwarzschild BH, this value of ISCO corresponds to a mass of the compact object $M_1 \approx c^2R_{\rm ISCO}/6G \approx (5.7 \pm 0.7)\,d_{10}\,M_{\odot}$; for a moderate spin parameter $a_{\ast} \approx 0.3$, $M_1 \approx c^2R_{\rm ISCO}/5G \approx (6.9 \pm 0.8)\,d_{10}\,M_{\odot}$;
for a maximally spinning Kerr BH, $M_1 \approx c^2R_{\rm ISCO}/G \approx (34 \pm 4)\,d_{10}\,M_{\odot}$. The inner-disk constraint to the BH mass is $M_1 > 5.0\,d_{10}\,M_{\odot}$ (approximately equivalent to a 90\% confidence limit).

\begin{table}
\caption{Main properties of MAXI J0637$-$430}
%\vspace{-0.4cm}
\begin{center}
\begin{tabular}{lr}  
\hline \hline\\[-7pt]    
Quantity & Origin\\
\hline
\multicolumn{2}{c}{Values taken from published literature}\\
\hline\\[-5pt]
$r_{\rm in} \sqrt{\cos \theta} = (32 \pm 1)\,d_{10}$ km & [1,2]\\[3pt]
$R_{\rm out} = (6.1 \pm 0.7) \times 10^{5} d_{10}$ km & [1] \\[3pt]
$2V \sin \theta \approx (2000 \pm 100)$ km s$^{-1}$ & [1]\\[3pt]
$\theta = 64^{\circ} \pm 6^{\circ}$  & [3]\\[3pt]
$\tau = (53 \pm 1)$ d   & [1]\\[3pt]
$\alpha = 0.34^{+0.06}_{-0.05}$    & [1] \\[3pt]
$L_{\rm peak,obs} \approx (1.5 \pm 0.3) \times 10^{38} \left(d_{10}\right)^2$ erg s$^{-1}$ & [1,2]\\[3pt]
\hline  \\[-7pt]
\multicolumn{2}{c}{Main values or constraints derived in this paper}\\
\hline\\[-5pt]
$R_{\rm ISCO} \approx (51 \pm 6) \, d_{10}$ km   & [4]  \\[3pt]
$M_1 > 5.0 \, d_{10} M_{\odot}$   & [4]  \\[3pt]
$M_1 < 6.8 \, d_{10} M_{\odot}$   & [5] \\[3pt]
$d \approx (8.7 \pm 2.3)\,(\eta/0.1)^{-9/17}\ \ {\mathrm{kpc}}$ & [6]\\[3pt]
$|z| \approx (3.1 \pm 0.8)
%\,(\eta/0.1)^{-9/17}
\ \ {\mathrm{kpc}}$  & [7]\\[3pt]
$r \approx (13.6 \pm 1.9)$ kpc  & [7]\\[3pt]
$R \approx (13.2 \pm 1.8)$ kpc & [7]\\[3pt]
%$R \approx 13.2^{+1.8}_{-1.7}$ kpc & [7]\\[3pt]
$M_1 \approx (5.1 \pm 1.6)\,(\eta/0.1)^{-9/17}\,M_{\odot}$ & [8]\\[3pt]
$\alpha_{\ast} \lesssim 0.25$ & [9]\\[3pt]
$L_{\rm peak} \approx (1.1\pm0.6)\,(\eta/0.1)^{-18/17}\, \times 10^{38}\ \ {\mathrm{erg~s}}^{-1}$ & [10]\\[3pt]
$\lambda_{\rm peak} \approx (0.17\pm0.11)\,(\eta/0.1)^{-9/17}$ & [10]\\[3pt]
$q \approx 0.04$--0.06  & [11]\\[3pt] 
$M_2 \approx (0.25 \pm 0.07)M_{\odot}$ & [11]\\[3pt]
${P_{\rm orb}} \approx 2.2^{+0.8}_{-0.6}\ \ \mathrm{hr}$ & [11]\\[3pt]
\hline
\end{tabular} 
\label{tab2}
\end{center}
\vspace{-0.3cm}
\begin{flushleft} 
%References:\\
1: from \cite{tetarenko21};\\
2: from \cite{ma22};\\
3: from \cite{lazar21};\\
4: from the inner disk radius;\\
5: from the He {\footnotesize{II}} $\lambda 4686$ peak-to-peak separation;\\
6: equating $L_{\rm peak,obs}$ to the predicted value for a standard disk model;\\
7: from the source distance ($\eta = 0.1$ for simplicity) and coordinates, assuming a heliocentric distance of 8.2 kpc from the Galactic centre;\\
8: combining inner disk (X-ray) constraint, outer disk (optical) constraint, and distance estimate;\\
9: combining the best estimate of $M_1$ with the inner disk radius constraint;\\
10: from $L_{\rm peak,obs}$ combined with $d$ and $M_1$;\\
11: from Roche lobe geometry and period-density relation \citep{eggleton83}.
\end{flushleft}
\end{table}

\subsection{Mass constraints from the outer-disk emission}
By fitting the combined {\it Swift}/XRT and UVOT data with the irradiated disk model {\tt diskir}, \cite{tetarenko21} provided an estimate of the outer disk radius $R_{\rm out}$ throughout the outburst. Moreover, their optical spectra taken with the Gemini Multi-Object Spectrograph show double-peaked H and He emission lines of clear disk origin \citep{tetarenko21}. The observed peak separation $2V$ (averaged over the orbital phase) of a disk emission line is approximately twice the projected rotational velocity of the outermost ring (at radius $R^{\prime}_{\rm out}$) that emits that particular line \citep{smak81,paczynski77,horne86,warner95}. This provides a powerful constraint on the BH mass.

We use the He {\footnotesize II} $\lambda 4686$ line profile (as published in \citealt{tetarenko21}) for our estimates. \cite{tetarenko21} modelled the emission lines with double Gaussians. For a disk line, the velocity separation between the two fitting Gaussians is always slightly smaller than the peak separation in the definition of \cite{smak81}. However, we can estimate the He {\footnotesize II} $\lambda 4686$ peak separation directly from Fig.~6 of \cite{tetarenko21} as $2V \sin \theta \approx (2000 \pm 100)$ km s$^{-1}$ on MJD 58842 (the Gemini observation with the highest signal-to-noise ratio, in which the sharp peaks clearly stand out). Then, we can apply the relation $V \approx \left(GM_1/R^{\prime}_{\rm out}\right)^{1/2}$, and determine $M_1$ if we can determine $R^{\prime}_{\rm out}$.

The simplest assumption at this point is that for He {\footnotesize II} $\lambda 4686$ emission, $R^{\prime}_{\rm out} = R_{\rm out}$ (the latter determined with {\tt diskir}). 
%How realistic is this assumption? 
If the outer annuli of the disk are too cold to emit He {\footnotesize II} $\lambda 4686$, $R^{\prime}_{\rm out} < R_{\rm out}$, and we would be over-estimating the BH mass. However, in MAXI J0637$-$430, the disk is relatively compact (as we will mention later). Based on continuum modelling, the effective temperature at the outer edge of the disk is $\approx$25,000 K on MJD 58842 \citep{tetarenko21}. More importantly, what determines the line emission from an accretion disk in the hot state is not the effective temperature but the presence of an optically thin temperature inversion layer (hotter than the effective temperature at the same radius), or chromosphere, at the disk surface \citep{smak91,sakhibullin98,wu01,wickramasinghe05}. This chromosphere is generated by the X-ray irradiation, particularly of soft X-ray photons from the inner disk (abundant in the high/soft state). Thus, we argue that the irradiated surface of the outer disk can produce He {\footnotesize II} $\lambda 4686$ and our assumption of $R^{\prime}_{\rm out} = R_{\rm out}$ is acceptable.

Next, we need to determine $R_{\rm out}$. The maximum extension of the outer disk was also reached just after outburst peak (MJD 58796.74, {\it Swift}/XRT ObsID 00012172003), with $R_{\rm out} \approx 10^{4.10\pm0.03}\,r_{\rm in} \approx (6.1 \pm 0.7) \times 10^{10}\,d_{10}$ cm \citep{tetarenko21}. It remained essentially unchanged within the errors for the following 10 days (MJD 58798--58807), with $R_{\rm out} \approx (5.8 \pm 0.7) \times 10^{10}\,d_{10}$ cm on MJD 58807.76 (ObsID 12172013). By the time of the third Gemini observation on MJD 58842 (2019 December 25), it had declined to $R_{\rm out} \approx (3.5 \pm 0.5) \times 10^{10}\,d_{10}$ cm \citep{tetarenko21}. The decline seen from a {\tt diskir} fit to the continuum optical emission is the result of the outer annuli either becoming cold and optically thin, or disappearing altogether from the inflow. However, X-ray irradiated, optically thin gas in the outer disk would still be able to emit optical lines with a double-peaked profile (by comparison, A0620$-$00 is a classic example of Galactic BH with double-peaked optical emission lines even in a very low X-ray state: \citealt{johnston89,orosz94}). From the peak separation relation $M_1 \approx V^2 \, R_{\rm out}/G$, we obtain $M_1 \approx (5.7 \pm 1.1)\, d_{10} M_{\odot}$ if we assume the maximum observed outer disk, or $M_1 \approx (3.3\pm0.7)\, d_{10} M_{\odot}$ if the line emission came only from the optically thick portion of the disk on MJD 58842.
%$M_1 \approx 5.66\, d_{10} M_{\odot}$ if we assume the maximum observed outer disk, or $M_1 \approx 3.25\, d_{10} M_{\odot}$ if the line emission came only from the optically thick portion of the disk on MJD 58842.
%https://duetosymmetry.com/tool/kerr-calculator-v2/

There are at least two other possible sources on uncertainty on $V$ we need to be aware of. The projected velocity of the disk edges (hence, the observed peak separation) oscillates sinusoidally with a period equal to half the binary period \citep{paczynski77,liu20}, because the outer disk is not perfectly circular; the variability $\Delta V/V$ can be as large as $\sim$10 per cent. We do not know what phases of the orbital cycle the system was observed in, when the Gemini observations were taken. However, the log of the Gemini observations for MJD 58842 \citep{tetarenko21} shows a series of evenly spaced 600-s exposures, covering about 3.5 hr. We will show later (Section 4) that this is enough to fully cover at least one binary period. Thus, the error on the mass estimate due to the $\Delta V/V$ oscillation is substantially less than 10 per cent and is negligible compared with other sources of error. The second effect that may 
alter the apparent peak separation is the presence of an outflow. If a line is emitted from a disk wind rather than from the thin disk chromosphere, its peak separation and full-width half maximum decrease, and the peaks become more smoothed \citep{matthews15,murray96,murray97}. This would cause an under-estimate of the BH mass derived from peak separation. However, the sharpness of the Smak-type line profile \citep{smak81} observed on MJD 58842 suggests that this effect is not significant here. 

In summary, the outer disk constraints provide a robust upper limit $M_1 < 6.8\, d_{10} M_{\odot}$, approximately equivalent to a 90\% confidence limit. Its robustness is based on the plausible assumption (supported also by the {\tt diskir} modelling) that the outer disk during the Gemini observations is not larger than it was at outburst peak, and on the fact that the He {\footnotesize II} emitting region must be $\leq R_{\rm out}$. The lower limit to the BH mass from the outer disk constraints is less robust and more model dependent, because it requires an estimate of the outermost disk annulus still emitting He {\footnotesize II} $\lambda$4686 at the epoch of the optical spectra. However, luckily, an accurate knowledge of the lower mass limit from the optical lines is not required for our BH mass estimate, because a more reliable lower limit is already provided by the X-ray continuum fitting (Section 2.1).

\subsection{Combining inner and outer disk constraints}
The most likely range for the BH mass is where the estimate from the inner disk (Section 2.1) and from the outer disk (Section 2.2) overlap. This happens for $5.0\,d_{10} < M_1/M_{\odot} < 6.8\, d_{10}$ (approximately a 90\% confidence limit). Henceforth, for simplicity, we will take the half point of this range and use $M_1 \approx (5.9 \pm 0.9)\, d_{10} M_{\odot}$ as our best mass estimate. The best-fitting value for a Schwarzschild BH ($M_1 \approx 5.7 \, d_{10} M_{\odot}$) is very close to the central value of our estimate; higher spins correspond to higher BH masses. The 90\% upper mass limit $M_1 \approx 6.8 \, d_{10} M_{\odot}$ corresponds to $a_{\ast} \approx 0.25$.

%should be consistent with each other. The only range of BH masses where some estimates from the two methods overlap is for $M_1 \approx 5.0\,d_7\,M_{\odot}$ (more accurately $M_1 \approx 4.7\,d_7\,M_{\odot}$). We will adopt the latter value for the next steps in our estimate of the system parameters.

\section{Distance and Eddington ratio}
The next step is to determine the distance, which we have kept as a free parameter so far. For this, we use two empirical results not exploited so far in our analysis: the peak luminosity and the exponential decay timescale.

The peak outburst luminosity ($L_{\rm peak}$) in a transient low-mass X-ray binary is proportional to the peak accretion rate ($\dot{M}_{\rm peak}$): $L_{\rm peak} \approx \eta \dot{M}_{\rm peak} c^2\approx 0.1 \dot{M}_{\rm peak} c^2$, where $\eta$ is the radiative efficiency. From \cite{king98}, 
\begin{equation}
\dot{M}_{\rm peak} \approx R_{\rm out}\,\nu\,\rho,
\end{equation}
where $\nu$ is the kinematic viscosity and $\rho$ is the mass density. Viscosity and density obviously vary as a function of radius in the disk, but this approximation is a good match to observed light-curves of BH transients (see also \citealt{powell07}) when we use the values of $\nu$ and $\rho$ in the outer disk, which contains most of the mass. After the peak, the emitted luminosity declines on an exponential timescale $\tau \approx \left(R_{\rm out}\right)^2/3\nu$ \citep{king98}. 

For the 2019 outburst of MAXI J0637$-$430, \cite{tetarenko21} determined an exponential decay timescale $\tau \approx (53 \pm 1)$ d. We have already estimated (Section 2.2) that just after outburst peak, $R_{\rm out} \approx (6.1 \pm 0.7) \times 10^{10}\,d_{10}$ cm.
From this, we obtain a value of $R_{\rm out}\,\nu = \left(R_{\rm out}\right)^3/(3\tau) \approx (16.5 \pm 5.7) \times 10^{24} \left(d_{10}\right)^3$ cm$^3$ s$^{-1}$. The density $\rho$ can be expressed for a standard disk as:
\begin{equation}
    \rho \approx 3.1 \times 10^{-8} \, \alpha^{-7/10} \, \dot{M}_{16}^{11/20} \, \left(m_1\right)^{5/8} \, R_{10}^{-15/8}\ \ {\mathrm{g~cm}}^{-3}
\end{equation}
\citep{ss73,frank02}, where $\alpha$ is the viscosity parameter, $\dot{M}_{16} \equiv \left(\dot{M}/10^{16}{\mathrm{~g~s}}^{-1}\right)$,
$m_1 \equiv M_1/M_{\odot} \approx (5.9 \pm 0.9)\,d_{10}$, $R_{10} \equiv \left(R_{\rm out}/10^{10}{\mathrm{~cm}}\right) \approx (6.1 \pm 0.7) \,d_{10}$. From their X-ray light-curve modelling, \cite{tetarenko21} estimated $\alpha \approx 0.34^{+0.06}_{-0.05}$, a value roughly in the middle of the distribution of viscosity parameters in BH transients \citep{tetarenko18}. 

Inserting $R_{\rm out}\,\nu$ and $\rho$ into Equation (1), and solving for $\dot{M}_{\rm peak}$, we obtain after straightforward algebra:
%\begin{eqnarray}
\begin{equation}
\dot{M}_{\rm peak}  \approx (2.1\pm 0.9) \times 10^{18}\,\left(d_{10}\right)^{35/9}\ \ {\mathrm{g~s}}^{-1}
\end{equation}
For the conversion from model-predicted accretion rate to model-predicted luminosity, we need to make an assumption on the efficiency $\eta$, for which we have no direct empirical constraints. For convenience of comparison with other studies of BH transients, we will normalize our results as a function of the parameter $\eta = 0.1$ but we will explicitly include the dependence on $(\eta/0.1)$. The low spin of the BH in MAXI J0637$-$430 and the possible energy release through disk outflows are two arguments in support of an efficiency slightly lower than 0.1 (probably $0.06 \lesssim \eta \lesssim 0.1)$. Hence, we obtain:
\begin{equation}
L_{\rm peak,mod} \approx (1.9\pm0.8) \times 10^{38}\,(\eta/0.1)\,\left(d_{10}\right)^{35/9}\ \ {\mathrm{erg~s}}^{-1}.
\end{equation}

The unabsorbed bolometric flux at peak outburst, derived from multi-instrument broadband observations \citep{ma22,tetarenko21}, is $F_{\rm bol} \approx (1.1 \pm 0.2) \times 10^{-8}$ erg cm$^{-2}$ s$^{-1}$, where the uncertainty takes also into account the different set of best-fitting values across different fitting models ({\it e.g.}, disk-blackbody plus power-law, or Comptonization models). For X-ray binaries, the conversion between fluxes and luminosities is generally taken to follow the scaling $L \approx 2\pi/(\cos \theta)\,F\,d^2$ for the disk emission component, and $L \approx 4\pi\,F\,d^2$ for the Comptonized component. This introduces an additional model dependency on how to separate the disk-like and isotropic flux contributions. However, luckily, for a viewing angle $\theta = 64^{\circ} \pm 6^{\circ}$, the two scaling relations are almost identical. Including this uncertainty, we estimate the peak unabsorbed bolometric luminosity as: 
\begin{equation}
L_{\rm peak,obs} \approx (1.5\pm0.3) \times 10^{38}\,\left(d_{10}\right)^{2}\ \ {\mathrm{erg~s}}^{-1}.
\end{equation}

Equating the model and empirical luminosities in Equations (4,5), we can now solve for the distance:
\begin{equation}
d \approx (8.7 \pm 2.3)\,(\eta/0.1)^{-9/17}\ \ {\mathrm{kpc}}
\end{equation}
and a vertical distance $|z|$ above the Galactic plane of 
\begin{equation}
|z| \approx (3.1 \pm 0.8)\,(\eta/0.1)^{-9/17}\ \ {\mathrm{kpc}}.
\end{equation}

Now, we can finally express the BH mass as:
\begin{equation}
M_1 \approx (5.1 \pm 1.6)\,(\eta/0.1)^{-9/17}\,M_{\odot},
\end{equation}
the peak bolometric luminosity as 
\begin{equation}
L_{\rm peak} \approx (1.1\pm0.6)\,(\eta/0.1)^{-18/17}\, \times 10^{38}\ \ {\mathrm{erg~s}}^{-1},
\end{equation}
and the Eddington ratio $\lambda_{\rm peak} = L_{\rm peak}/L_{\rm Edd}$ as:
\begin{equation}
\lambda_{\rm peak} \approx (0.17\pm0.11)\,(\eta/0.1)^{-9/17}.
\end{equation}

%In summary, the peak luminosity of MAXI J0637$-$430 is $L_{\rm peak} \approx 1 \times 10^{38}$ erg s$^{-1}$. For a mass $M_1 \approx 5\,M_{\odot}$, the Eddington luminosity $L_{\rm Edd} \approx 6.5 \times 10^{38}$ erg s${-1}$. Thus, the outburst may have peaked at $\approx$0.15 times the Eddington luminosity. The transition to the hard state during the decline happened at $\approx$0.010 $L_{\rm Edd}$ \citep{tetarenko21}. Both the peak luminosity and the transition back to the hard state are in the expected range for systems with such a short binary period \citep{maccarone03,tetarenko16}.
% Knevitt, G. ; Wynn, G. A. ; Vaughan, S. ; Watson, M. G. 2014MNRAS.437.3087
%  Arur, K. ; Maccarone, T. J. 2018MNRAS.474...69
%   Tetarenko, B. E. search by orcid ; Sivakoff, G. R. search by orcid ; Heinke, C. O. search by orcid ; Gladstone, J. C.  2016ApJS..222...15

\section{Mass of the donor star and binary period}

Our next objective is an estimate of the mass ratio (defined here as $q \equiv M_2/M_1$) and binary period. For this, we use the well-known Roche lobe approximations of \cite{eggleton83}:
\begin{equation}
R_{\rm RL2} \approx a\, \left[\frac{0.49\,q^{2/3}}{0.60\,q^{2/3} + \ln \left(1+q^{1/3}\right)}\right],
\end{equation}
where $R_{\rm RL2}$ is the radius of the donor star's Roche lobe. We plausibly assume that the donor star is filling its Roche lobe, so that its radius $R_2 = R_{\rm RL2}$. The expression for $R_{\rm RL1}$ (Roche lobe of the BH) is identical to Equation (10), with $q \rightarrow q^{-1}$:
\begin{equation}
R_{\rm RL1} \approx a\, \left[\frac{0.49\,q^{-2/3}}{0.60\,q^{-2/3} + \ln \left(1+q^{-1/3}\right)}\right].
\end{equation}

Equations (10,11) can be solved for $a$ and $q$ if we have two empirical constraints on $R_{\rm RL1}$ and $R_{\rm RL2}$. The first constraint is that the maximum size of the outer disk (tidal truncation radius) is $R_{\rm out} \approx 0.8 R_{\rm RL1}$, corresponding to the 3:2 resonance radius \citep{whitehurst91,eggleton83,paczynski77}. Hence, 
\begin{eqnarray}
R_{\rm RL1} & \approx & (6.1 \pm 0.7)/0.8 \times 10^{10}\,d_{10}\ \ {\rm cm} \nonumber \\
& \approx & (6.6\pm1.9)\,(\eta/0.1)^{-9/17}\,10^{10}\ \ {\rm cm} \nonumber\\
&\approx & (0.95\pm0.27)\,(\eta/0.1)^{-9/17} R_{\odot}.
\end{eqnarray}
The second constraint is that the donor star is a Roche-lobe-filling low-mass star. Such stars have a main-sequence life span much longer than the Hubble time: therefore, we can take as a plausible assumption that the donor is still on main-sequence. For main-sequence dwarfs with masses between $\approx$0.1--1$M_{\odot}$, both the theoretical mass-radius relation and empirical observations suggest that $M_2/M_{\odot} \approx R_2/R_{\odot}$ \citep{baraffe15,parsons18}, and $R_2/R_{\odot} \approx R_{\rm RL2}/R_{\odot}$ for mass transfer to occur.
%(an assumption well justified a posteriori), 
%so that $M_2/M_{\odot} \approx R_2/R_{\odot} = R_{\rm RL2}/R_{\odot}$. 
Hence, 
\begin{equation}
R_{\rm RL2}/R_{\odot} \approx q\,M_1/M_{\odot} \approx (5.1 \pm 1.6)\,q\,(\eta/0.1)^{-9/17}.
\end{equation}

%, given the low mass range that we will find from this calculation), so that $M_2/M_{\odot} \approx R_2/R_{\odot}$.
%size of the Roche lobe for the primary ($R_{\rm RL1}$). We do not have a constraint on the mass ratio $q \equiv M_2/M_1$ yet. However, a convenient and widely used approximation for the maximum outer disk size is $R_{\rm out} \approx 0.8 R_{\rm RL1}$. Hence, $R_{\rm RL1} \approx 0.91\,d_7\,R_{\odot}$. Let us combine this expression with another useful approximation for the secondary Roche lobe radius, $R_{\rm RL2} \approx R_{\rm RL1} \,q^{0.45}$, valid for $0.03\lesssim q \lesssim 1$ \citep{frank02}. Moreover, we plausibly assume that the donor star is a low-mass star filling its Roche lobe, so that the radius $R_2$ if the donor star is $R_2 = R_{\rm RL2}$. Finally, let us assume that the star is on the main sequence (an assumption well justified a posteriori, given the small mass that we will find from this calculation), so that $M_2/M_{\odot} \approx R_2/R_{\odot}$.

%Solving the previous equations for $M_2$, we obtain 
%$M_2/M_{\odot} \approx  \left(R_{\rm RL1}/R_{\odot}\right)^{1/0.55}  \left(M_1/M_{\odot}\right)^{-0.45/0.55}$. Finally, $M_2 \approx 0.24\,d_7\,M_{\odot}$ and $q \approx 0.05$.

After inserting the empirical values of $R_{\rm RL1}$ and $R_{\rm RL2}$ into Equations (11,12), and taking the ratio of the two expressions, we can solve numerically for $q$. For $d \approx 8.7$ kpc, $M_1 \approx 5.1 M_{\odot}$ and $\eta \approx 0.1$, we obtain $q \approx 0.049$, corresponding to $M_2 \approx 0.25 M_{\odot}$. We also calculated a grid of numerical solutions over the acceptable ranges of $d$, $M_1$ and $\eta$, and found that in all cases, $q \approx 0.04$--0.06 and $M_2 \approx (0.25 \pm 0.07)M_{\odot}$. (Error ranges for the various quantities are not independent, so a simple Gaussian propagation is not appropriate here). As a general trend, larger distances (for all other parameters fixed) correspond to higher values of both $M_1$ and $M_2$, with no significant change to $q$. The same thing happens if we reduce the radiative efficiency. The binary separation is $a \approx (1.5 \pm 0.4) (\eta/0.1)^{-9/17} R_{\odot} \approx (1.1 \pm 0.3) (\eta/0.1)^{-9/17} 10^{11}$ cm.

%We can recover a similar result from a more refined approximation of the Roche lobe quantities. From the tidal truncation radius, we derive the binary separation $a \approx \left(R_{\rm out}/0.60\right)\,(1+q)$ \citep{warner95}. From the approximation of \cite{eggleton83},
%\begin{equation}
%\left(\frac{M_2}{M_{\odot}}\right) \approx \left(\frac{R_{\rm RL2}}{R_{\odot}}\right) \approx \frac{1+q}{0.60} \left(\frac{R_{\rm out}}{R_{\odot}}\right) \left[\frac{0.49\,q^{2/3}}{0.60\,q^{2/3} + \ln \left(1+q^{1/3}\right)}\right].
%\end{equation}
%Then, substituting into (1) our adopted values of $M_1 \approx 4.7\,d_7\,M_{\odot}$ and $R_{\rm out} \approx 0.73\,d_7\,R_{\odot}$,
%\begin{equation}
%q \approx 0.26\,(1+q) \left[\frac{0.49\,q^{2/3}}{0.60\,q^{2/3} + \ln \left(1+q^{1/3}\right)}\right].
%\end{equation}
%Solving numerically for $q$, we obtain $q \approx 0.044$ and $M_2 \approx 0.21\,d_7\,M_{\odot}$.

Finally, our estimate of $q$ enables us to constrain the binary period $P_{\rm orb}$, via the period-density relation \citep{eggleton83,frank02}. We assume again that the donor star is a main-sequence dwarf filling its Roche lobe, with average density $\rho_2 \approx 1.4\,\left(M_2/M_{\odot}\right)^{-2}$ g cm$^{-3}$. Then, 
\begin{equation}
\left(\frac{P_{\rm orb}}{{\rm d}}\right)  \left(\frac{\rho_2}{{\rm{g~cm}^{-3}}}\right)^{1/2} \approx 0.1375 \left(\frac{q}{1+q}\right)^{1/2} \left(\frac{R_2}{a}\right)^{-3/2} \approx 0.434,
%\approx 2.5\,d_{10}\ \ \mathrm{hr}.
\end{equation}
and therefore, $P_{\rm orb} \approx 2.2$ hr. To determine the uncertainty range, we determined the numerical solutions for $q$ and $M_2$ over the plausible range of $d$, $M_1$ and $\eta$. We estimate that
\begin{equation}
{P_{\rm orb}} \approx 2.2^{+0.8}_{-0.6}\ \ \mathrm{hr}
\end{equation}

An alternative, simpler way to estimate the period is to use the approximate relation $R_{\rm RL2}/R_{\rm RL1} \approx q^{0.45}$ \citep{frank02}, instead of the ratio between Equations (11) and (12). Then, inserting the values of $R_{\rm RL2}$ and $R_{\rm RL1}$ determined earlier, we can write
\begin{equation}
q \approx \left(\frac{0.95\pm0.27}{5.1\pm1.6}\right)^{1/0.55} \approx 0.047 \pm 0.016.
\end{equation}
For the estimate of the uncertainty range in Equation (17), we cannot simply propagate the errors at the numerator and denominator, because they are correlated via the same distance factor. Instead, we separated the error contribution of the distance factor (a term that cancels out at numerator and denominator) and of the other factors (related to the estimate of inner and outer disk radii). Only the latter terms contribute to the error in $q$.

From $M_1$ and $q$, we then obtain $M_2 \approx (0.24 \pm 0.11)M_{\odot}$. Finally, from the period-mass relation \citep{frank02}, 
\begin{equation}
{P_{\rm orb}} \approx (1/0.11)\,\left(M_2/M_{\odot}\right) \ \mathrm{hr} \approx (2.2 \pm 1.0)\ \ \mathrm{hr},
\end{equation}
perfectly consistent with the first derivation (Equation 15) based on the Eggleton approximation.

Finally, let us consider the possibility that the secondary star deviates from the approximate relation $M_2/M_{\odot} \approx R_2/R_{\odot}$, and what effect this may have on our estimate of the binary period. One situation where this happens is in close binary systems with a compact object and an M star, because of the fast rotation of the star, even when it is detached from its Roche lobe (no mass transfer). In particular, \cite{parsons18} showed that the radii of M dwarfs in short-period binaries with a white dwarf are up to $\sim$10 per cent larger than the theoretical radii of isolated stars.  To estimate whether this discrepancy can significantly affect our estimate of the binary period in MAXI J0637$-$430, we repeated the previous derivation (from the ratio of Equations 11,12) but with the revised conditions that $R_{\rm RL2}/R_{\odot} \approx 1.1 q\,M_1/M_{\odot}$ and $\rho_2 \approx 1.4\times (1.1)^{-3}\,\left(M_2/M_{\odot}\right)^{-2}$ g cm$^{-3}$. We recover the same period $P_{\rm orb} \approx 2.2^{+0.8}_{-0.6}$ hr derived earlier, but with a slightly smaller donor mass $M_2 \approx (0.21 \pm 0.07)M_{\odot}$ ($q \approx 0.03$--0.05). Another situation where the M-type secondary may be over-inflated compared with an isolated main-sequence star is when there is mass transfer in the system. If the mass-transfer time-scale is shorter than the donor's thermal time-scale, the secondary will not be able to maintain thermal equilibrium and will expand, compared with an isolated main-sequence star. This effect has been studied \citep{patterson05,knigge06} in cataclysmic variables ({\it i.e.}, compact binaries with an accreting white dwarf). As a simple test, we repeated our derivation of the period in MAXI J0637$-$430 assuming that the mass-radius relation is the same as the empirical relation of \cite{knigge06} (see their Equation 9 and Fig.~3). We obtain a best-fitting period of $\approx$2.1 hr, and a donor mass $M_2 \approx 0.20 M_{\odot}$.

\section{Optical counterpart}

Having estimated both $M_1$ and $M_2$, we can also predict the projected radial semi-amplitude $K_1$ of the primary in its orbital motion:
\begin{equation}
    K_1 = \left[\frac{2\pi G\,M_2^3\,\left(\sin \theta\right)^3}{P_{\rm orb}\,\left(M_1+M_2\right)^2}\right]^{1/3} \approx (34.7\pm7.6)\ \ {\mathrm{km~s}}^{-1}.
\end{equation}
Although $M_1$ and $M_2$ contain a distance factor (and its relative uncertainty), $K_1$ does not depend on the distance, at least as a first approximation. That is because $P_{\rm orb} \propto M_2$ (Equation 17), hence $K_1 \propto [q/(1+q)]^{2/3} \sin \theta$. The projected radial semi-amplitude $K_2$ of the companion star is simply $q K_1 \approx 710$ km s$^{-1}$.
%independent of distance, because both masses and $P_{\rm orb}$ scale as $d_{10}$. The uncertainty quoted here comes simply from the 90\% confidence limit of $\theta$. 

The accretion disk follows the orbital motion of the BH. For this reason, the semi-amplitude $K_1$ is often used to describe also the sinusoidal motion of the central position of the disk emission lines. However, for a disk extended close to the Roche lobe, because of its tidal deformation, the observed velocity semi-amplitude of the central line position is a factor of $\approx$1.2 higher than the semi-amplitude of the BH motion \citep{huang67,paczynski77,liu20}. Hence, we expect $K_{\rm disk} \approx 40$ km s$^{-1}$ for the sinusoidal motion of the He {\footnotesize{II}} $\lambda 4686$ emission. This is well within the capability for phase-resolved spectroscopy of an 8-m class telescope when the system is in outburst. The major technical difficulty is of course that the binary period is only $\sim$2 hr.

Detecting the donor star in quiescence would be much more challenging. A main sequence star with a mass $\approx$0.25 $M_{\odot}$ has an M3.4--M4 type spectrum, with absolute brightness $M_V \approx 12.5$ mag
%, $M_V \sim 13$ mag, $M_R \sim 12$ mag, $M_I \sim 10$ mag 
\citep{pecaut13},
%\footnote{See also updated tables at http://www.pas.rochester.edu/\~emamajek/EEM\_dwarf\_UBVIJHK\_colors\_Teff.txt, by the same authors.}
to which we need to add a distance modulus of $\approx$14.7 mag (at a distance of 8.7 kpc), and an additional extinction term. The absorbing column density $N_{\rm H} \approx 2 \times 10^{21}$ cm$^{-2}$ found from X-ray spectral modelling \citep{ma22,tetarenko21,lazar21} corresponds to $A_V \approx 0.7$--1.0 mag \citep{foight16,willingale13,watson11,guver09}. However, X-ray estimates of the column density in X-ray binaries typically over-estimate the optical reddening. More plausibly, the reddening relevant to the optical counterpart is $E(B-V) \approx 0.065$--0.070 mag, corresponding to $A_V \approx 0.20$--0.25 mag \citep{tetarenko21}. Thus, we expect that the quiescent counterpart is undetectable in the $V$ band ($V \sim 27.5$ mag), but should be detectable in the near-infrared, at $I \sim 25$ mag, $K_{\rm s} \approx 22$ mag \citep{pecaut13,knigge06}.

An alternative (hypothetical) possibility is to measure the radial velocity of the BH from the Doppler shift of its X-ray absorption lines from an accretion disk wind, the next time that the system is observed in outburst. This is a technique proposed by \cite{zhang12} and already successfully applied to a few BH and neutron star X-ray binaries \citep{zhang12,ponti18}. The next generation of X-ray observatories ({\it e.g., XRISM, Athena}) with high-resolution spectrographs will be able to measure such orbital motions easily, if the X-ray spectrum does show such lines. The recurrence timescale for full outbursts in MAXI J0637$-$430 may be several decades (see also Section 6), but this technique may be useful for other short-period BH candidates similar to MAXI J0637$-$430, undetected in outburst so far.

Finally, the observed optical brightness in outburst, peaking at $V \approx 16.3$ mag \citep{tetarenko21} provides an additional, simple test that our distance and BH mass estimates are at least self-consistent and within the ballpark. For this, we use the well-known empirical relation of \cite{van94} (based on the size of the irradiated disk): 
\begin{equation}
    M_V = 1.57 (\pm 0.24) - 2.27 (\pm 0.32) \log \left[\left(P_{\rm orb}/1{\rm hr}\right)^{2/3} \lambda^{1/2}  \right]
\end{equation}
Assuming our best-estimate values of $P_{\rm orb} \approx 2.2$ hr and peak Eddington ratio $\lambda \approx 0.17$, the expected absolute visual magnitude is $M_V \approx (1.9 \pm 0.5)$ mag. At a distance of 8.7 kpc and with an additional 0.2 mag of extinction, this corresponds to $V \approx (16.8 \pm 0.5)$ mag, in reasonable agreement with the data.

\section{Discussion and Conclusions}
Using the X-ray and optical results available in the literature, we showed (Table 1) that MAXI J0637$-$430 is consistent with a low-mass BH ($M_1 \sim 4$--7 $M_{\odot}$) fed by an M-dwarf donor star with $M_2 \sim 0.2$--0.3 $M_{\odot}$. We also estimated a heliocentric distance $d \approx (8.7\pm2.3)$ kpc. The relatively low mass of the BH and its large distance imply that MAXI J0637$-$430 probably just exceeded the threshold luminosity of $\sim$0.1 $L_{\rm Edd}$ ($\approx$10$^{38}$ erg s$^{-1}$) in its 2019 outburst, enough to reach the thermal dominant state. The reason its outburst peaked at a low Eddington ratio compared with the majority of stellar-mass BH transients is the small size of the accretion disk (consequence of the small binary separation and short orbital period).

Previous work \citep{ma22} had assumed a heliocentric distance $d \lesssim 7$ kpc (vertical distance $\lesssim$2.5 kpc above the Galactic plane) to guarantee that the system belongs to either the thin or thick disk. For the distance estimate in this work, we relaxed this condition, allowing for a halo object. We assumed instead that its observed peak luminosity agrees with the predicted outburst peak luminosity in the thermal-viscous instability model. This gives our slightly larger best-estimate of $d \approx (8.7\pm2.3)$ kpc, corresponding to a vertical distance $|z| \approx (3.1 \pm 0.8)$ kpc, consistent with either the halo or the outer edge of the thick disk. Furthermore, MAXI J0637$-$430 is located away from the Galactic Centre ($l = 251^{\circ}.5320370, b = -20^{\circ}.67473903$). We estimate a spherical radial distance $r \approx (13.6 \pm 1.9)$ kpc from the Galactic Center\footnote{$r = \sqrt{\left(d \cos b \cos l - R_0\right)^2 + \left(d \cos b \sin l \right)^2 + \left(d \sin b\right)^2}$, and we adopt a  heliocentric distance from the Galactic Centre $R_0 = 8.2$ kpc \citep{gravity19}.}, and a cylindrical radial distance 
%$R \approx 13.2^{+1.8}_{-1.7}$ kpc. 
$R \approx (13.2 \pm 1.8)$ kpc. 
This is a very unusual and therefore interesting location for Galactic BH candidates. Only 3 out of 68 BH candidates listed (as of 2022 June) in the BlackCAT catalogue \citep{Corral-Santana2016}  have a Galactic latitude $|b| > 20^{\circ}$ and only 5 out of 68 have a Galactic longitude of $90^{\circ} < l < 270^{\circ}$. MAXI J0637$-$430 is the farthest source from the Galactic Centre among all currently known Galactic BH candidates.

%r = \sqrt{\left(d \cos b \cos l) - R_0\right)^2 + \left(d \cos b sin l \right)^2 + \left(d \sin b\right)^2}

The Milky Way gravitational potential is weaker (in absolute value) at larger distances: therefore, we expect that X-ray binaries born in the thin disk will have a higher root-mean-square value of $z$ for the same natal kick velocity \citep{repetto17}.  For example, the predicted $|z_{\rm rms}|$ at $R \approx 13$ kpc is twice as high as the predicted scale-height at $R \approx 6$ kpc \citep[][their Fig.~9]{repetto17}. This may partly explain the high Galactic latitude of MAXI J0637$-$430. However, even at $R \approx 13$ kpc, a height of $|z| \approx 3$ kpc is an outlier. Another factor may be at play, as we discuss next.

The other unusual property of MAXI J0637$-$430 is its very short binary period. \cite{tetarenko21} suggested $P_{\rm orb} \sim 2$--4 hr, based on its small outer disk size and low peak luminosity. When we include the additional observational and modelling constraints discussed in this work, we obtain a best-fitting value of $P_{\rm orb} \sim 2.2$ hr (90\% confidence range of $P_{\rm orb} \sim 1.5$--3 hr).
A period of 2.2 hr would be the shortest ever found in a Galactic BH transient \citep{Corral-Santana2016}. More importantly, it would be the first BH X-ray binary either inside of below the so-called ``period gap'', at $2.1 \lesssim P_{\rm orb}({\rm hr}) \lesssim 3.1$ \citep{spruit83,rappaport83,patterson84,king88,podsiadlowski02,webbink02,knigge06}.

The most common explanation for the relative lack of cataclysmic variables and X-ray binaries in this period range is that for masses $M_2 \lesssim 0.25 M_{\odot}$, the donor star becomes fully convective, and magnetic breaking stops or becomes much less efficient. The star contracts and (in most cases) loses contact with its Roche lobe, decreasing the mass transfer rate by orders of magnitude. From that point, the binary separation and the period continue to decrease slowly, only via gravitational wave emission. At some point around $P_{\rm orb} \sim 2$ hr, the donor star regains contact with its Roche lobe and mass transfer through the L1 Lagrangian point starts again, at a rate of $\sim$10$^{-10}$ $M_{\odot}$ yr$^{-1}$ \citep{king88,podsiadlowski02,knigge06}.  
%A detailed discussion of the predicted behaviour of X-ray binaries in or below the period gap is beyond the scope of this work. 
%This would place the system either in the period gap or just below it, in the regime of orbital shrinking driven by gravitational radiation.  
A period-gap location for MAXI J0637$-$430 would be problematic to explain, because the donor star is not expected to fill its Roche lobe in that evolutionary phase, and the mass transfer rate would be too low to trigger a disk-instability outburst. Instead, very faint X-ray transients are more likely to be in the period gap \citep{maccarone13}. Thus, we argue that MAXI J0637$-$430 is already below the period gap, when the donor star regains contact with its Roche lobe. If so, it would be the first example of BH X-ray binary with orbital shrinking driven by gravitational decay. Another consequence of such interpretation is that it should take $\sim$30 yr (a rough estimate within a factor of 2) to replenish the disk until the next outburst, considering that the total mass accreted during the 2019 outburst is $\sim$6 $\times 10^{24}$ g \citep{tetarenko21} and that the typical mass transfer rate driven by gravitational radiation corresponds to $\sim$2 $\times 10^{23}$ g per year.
%recurrence rate

However, much is still unknown about the new class of short-period BH X-ray binaries. There are now three Galactic BH candidates with a period inside or below the period gap: in addition to MAXI J0637$-$430, also MAXI J1659$-$152, with $P_{\rm orb} = (2.414 \pm 0.005)$ hr  \citep{kuulkers13,corral-santana18,torres21} and Swift J1357.2$-$0933, with $P_{\rm orb} = (2.8 \pm 0.3)$ hr \citep{charles19,mata15,armas13}. Both MAXI J1659$-$152 and Swift J1357.2$-$0933 reached peak X-ray luminosities much higher than expected if their donor star was not filling its Roche lobe. In particular, MAXI J1659$-$152 also reached the high/soft state (unlike Swift J1357.2$-$0933), with a peak luminosity $L_{\rm X} \sim 10^{38}$ erg s$^{-1}$ \citep{munoz-darias11,yamaoka12}, although its distance is also quite uncertain, with alternative estimates differing by at least a factor of 2 (see a discussion of this issue both in \citealt{kuulkers13} and \citealt{corral-santana18}). To solve this contradiction, \cite{kuulkers13} proposed that the donor star in MAXI J1659$-$152 is not a main-sequence M-type star, but is instead an evolved (He-rich), stripped star with an initial mass $\sim$1--1.5 $M_{\odot}$ (see also \citealt{pylyser88,pfahl03}), still filling its Roche lobe across what would have been the period gap. Whether the same explanation may apply to MAXI J0637$-$430 will be an interesting topic of future investigation. It is intriguing to notice the relative strength of the He {\footnotesize II} $\lambda 4686$ emission line, compared with the much weaker H {\footnotesize I} Balmer lines, in the Gemini spectra of \cite{tetarenko21}; however, further analysis of the optical line ratios is beyond the scope of this work. 
%In addition, the disk refilling time until its next outburst will constrain the long-term-average mass transfer rate and provide a useful test of the current evolutionary state of the system.

The most remarkable property shared by the three shortest-period BH candidates mentioned above is that they are all located at considerable height above the Galactic plane: $|z| \sim 2$ kpc for MAXI J1659$-$152 (considering again the large systematic uncertainties on its distance: \citealt{corral-santana18}) and $|z| > 4.6$ kpc for Swift J1357.2$-$0933 \citep{charles19}). A correlation between higher Galactic latitude and shorter binary periods for BH candidates has been noted and quantified before \citep{yamaoka12,kuulkers13,atri19,gandhi20} but is not entirely understood yet. An early explanation \citep{kuulkers13} was that BH binaries with lower total masses may receive a higher kick velocity (for the same linear momentum) when the BH is formed, and travel farther away from the Galactic plane; low-mass systems are also more likely to have a smaller donor star, lower binary separation and shorter orbital period. However, observations show no correlation between kick velocity and BH mass in X-ray binaries \citep[][their Fig.~10]{atri19}, contrary to this proposed scenario. A more plausible explanation \citep{repetto17,gandhi20} is based on the theoretical finding \citep{kalogera96,kalogera98,giacobbo20} that small kick velocities ($\lesssim$100 km s$^{-1}$) allow the post-supernova survival of wide binaries (long-period systems) as well as tight binaries (short-period systems); instead, kick velocities $>$100 km s$^{-1}$ strongly favour the survival of only short-period ones. This is because more compact systems have a higher binding energy and are less likely to be disrupted by a strong natal kick. Since the population of BH X-ray binaries found in the thick disk or halo is biased in favour of higher kick velocities ({\it i.e.}, they are mostly systems ejected from the thin disk rather than formed {\it in situ}), it will also be biased in favour of shorter periods. 
%(See also \citealt{repetto17} for a comparison of the predicted and observed spatial distributions and binary periods of Galactic BH and neutron star X-ray binaries).

In summary, the unassuming transient BH candidate MAXI J0637$-$430 may not be very impressive in terms of X-ray luminosity or other forms of activity, barely reaching $\sim$10\% of the Eddington luminosity at its peak, but it deserves further attention and in-depth studies (in particular, a deep near-infrared search of its quiescent counterpart) because it probes unexplored regions in the distribution of Galactic BH transients, and key aspects of binary evolution. It is probably the currently known Milky Way BH candidate most distant from the Galactic Centre; perhaps a halo object. It is probably also the BH X-ray binary with the shortest period, short enough to be driven by gravitational wave decay rather than magnetic breaking or nuclear evolution of the donor.

%BH transients with M3V--M4V donors with low peak luminosity in outburst, short binary periods, long duty cycles [search duty cycle in Knevitt paper] may represent an unexplored population of BH candidates in the Galactic halo. It may pose new challenges to theoretical models of BH X-ray binary formation, but also provide clues on the formation of thick disk and halo....blah blah...

%a very short binary period ($P_{\rm bin} \sim 2$ hr), located at a distance of $\sim$7 kpc. The shortest binary period found so far in Galactic BHs. Small binary separation, small disk, means low peak luminosity in the outburst. Enough to go into the high/soft state, although \cite{ma21} have suggested that the disk did not completely form. If such a small donor star and short period are confirmed, they would pose new challenges to theoretical models of BH X-ray binary formation. blah blah...

\section*{Acknowledgements}
This work is supported by the National Key R\&D Program of China (2021YFA0718500). 
RS acknowledges grant number 12073029 from the National Natural Science Foundation of China (NSFC). 
LT acknowledges funding support from NSFC under grant Nos. 12122306 and U1838115, the CAS Pioneer Hundred Talent Program Y8291130K2 and the scientific and technological innovation project of IHEP Y7515570U1.
We thank David Buckley, Rosanne Di Stefano, Jifeng Liu, Michela Mapelli, Thomas Russell, Ryan Urquhart and Yang Xie for useful discussions about Galactic X-ray binaries.
%This work is supported by the National Key R\&D Program of China (2021YFA0718500). We acknowledge funding support from the National Natural Science Foundation of China (NSFC) under grant Nos. U1838115, U1838201 and U1838202, the CAS Pioneer Hundred Talent Program Y8291130K2 and the Scientific and technological innovation project of IHEP Y7515570U.
We used Leo C.~Stein's ``Kerr ISCO calculator'' (https://duetosymmetry.com/tool/kerr-calculator-v2/) for our estimates of BH spin.  
%RS acknowledges hospitality at Sun Yat Sen University in Zhuhai, during part of this work. We thank Tom Jarrett for his explanations about star formation rates from {\it WISE} bands, and Alister Graham for comments about disk galaxy structure. We also appreciated the useful comments and suggestions from the referee. This publication makes use of data products from the Wide-field Infrared Survey Explorer, which is a joint project of the University of California, Los Angeles, and the Jet Propulsion Laboratory/California Institute of Technology, funded by the National Aeronautics and Space Administration. 

\section*{Data Availability}
The raw {\it Swift}/XRT and Gemini datasets used for this work are all available for download from their respective public archives. The {\it Swift}/XRT reduced data can be provided upon request. For the Gemini optical spectra, we used results already published in the literature.


\begin{thebibliography}{99}

\bibitem[Armas Padilla et al.(2013)]{armas13}  Armas Padilla M., Degenaar N., Russell D.M., Wijnands R., 2013, MNRAS, 428, 3083

\bibitem[Arnaud(1996)]{arnaud96} Arnaud K.A., 1996, ASPC, 101, 17

\bibitem[Arur \& Maccarone(2018)]{arur18} Arur K., Maccarone T.J., 2018, MNRAS, 474, 69

\bibitem[Atri et al.(2019)]{atri19} Atri P., et al. 2019, MNRAS, 489, 3116

\bibitem[Baraffe et al.(2015)]{baraffe15} Baraffe I., Homeier D., Allard F., Chabrier G., 2015, A\&A, 577, 42

\bibitem[\protect\citeauthoryear{Charles et al.}{2019}]{charles19}
 Charles P., Matthews J.H., Buckley D.A.H., Gandhi P., Kotze E., Paice J., 2019, MNRAS, 489, L47

\bibitem[\protect\citeauthoryear{Corral-Santana et al.}{2016}]{Corral-Santana2016} Corral-Santana J.M., Casares J., Mu{\~n}oz-Darias T., Bauer F.E., Mart{\'\i}nez-Pais I.G., Russell D.M., 2016, A\&A, 587, 61%. doi:10.1051/0004-6361/201527130

\bibitem[Corral-Santana et al.(2018)]{corral-santana18} Corral-Santana J.M., et al., 2018, MNRAS, 475, 1036

\bibitem[Davis \& Hubeny(2006)]{davis06} Davis S.W., Hubeny I., 2006, ApJS, 164, 530

\bibitem[Davis et al.(2005)]{davis05} Davis S.W., Blaes O.M., Hubeny I., Turner N.J., 2005, ApJ, 621, 372

\bibitem[Done \& Davis(2008)]{done08} Done C., Davis, S.W., 2008, ApJ, 683, 389

\bibitem[\protect\citeauthoryear{Dubus et al.}{1999}]{dubus99} Dubus G., Lasota J.-P., Hameury J.-M., Charles P., 1999, MNRAS, 303, 139 

\bibitem[Eggleton(1983)]{eggleton83} Eggleton P.P., 1983, ApJ, 268, 368

\bibitem[Foight et al.(2016)]{foight16} Foight D.R., G\"uver T., \"Ozel F., Slane P.O., 2016, ApJ, 826, 66

\bibitem[Frank et al.(2002)]{frank02} Frank J., King A.R., Raine D.J.,  Accretion Power in Astrophysics (Cambridge, UK: Cambridge University Press)

\bibitem[Gandhi et al.(2020)]{gandhi20} Gandhi P., Rao A., Charles P.A., Belczynski K., Maccarone T.J., Arur K., Corral-Santana J.M. 2020, MNRAS, 496, L.22

\bibitem[Giacobbo \& Mapelli(2020)]{giacobbo20} Giacobbo N., Mapelli M., 2020, ApJ, 891, 141 

\bibitem[Gierli\'nski \& Done(2004)]{gierlinski04} Gierli\'nski M., Done C.,  2004, MNRAS, 347, 885

\bibitem[Gravity Collaboration et al.(2019)]{gravity19} Gravity Collaboration, et al., 2019, A\&A, 625, L10

\bibitem[G\"uver \& \"Ozel(2009)]{guver09} G\"uver T., \"Ozel F., 2009, MNRAS, 400, 2050

\bibitem[Hameury \& Lasota(2020)]{hameury20} Hameury J.-M., Lasota J.-P., 2020, A\&A, 643, 171 

\bibitem[Horne \& Marsh(1986)]{horne86} Horne K., Marsh T.R., 1986, MNRAS, 218, 761

\bibitem[Huang(1967)]{huang67} Huang S.-S. 1967, ApJ, 148, 793

\bibitem[Jana et al.(2021)]{jana21} Jana A., Jaisawal G.K., Naik S.,  Kumari N., Chhotaray B., Altamirano D., Remillard R.A., Gendreau K.C., 
    2021, MNRAS, 504, 4793

\bibitem[Johnston et al.(1989)]{johnston89}  Johnston H.M., Kulkarni S.R., Oke J.B.,  1989, ApJ, 345, 492

\bibitem[Kalogera \& Webbink(1996)]{kalogera96} Kalogera V., Webbink R.F., 1996, ApJ, 458, 301

\bibitem[Kalogera \& Webbink(1998)]{kalogera98} Kalogera V., Webbink R.F., 1998, ApJ, 493, 351

\bibitem[Kennea et al.(2019)]{kennea19} Kennea J.A., et al., 2019, The Astronomer’s Telegram, 13257, 1
%et al.\ 2021%, arXiv:2104.13005

\bibitem[King(1988)]{king88} King A.R. 1988, QJRAS, 29, 1

\bibitem[King \& Ritter(1998)]{king98} King A.R., Ritter, H. 1998, MNRAS, 293, L42

\bibitem[Knevitt et al.(2014)]{knevitt14}  Knevitt G., Wynn G.A., Vaughan S., Watson M.G., 2014, MNRAS, 437, 3087

\bibitem[Knigge(2006)]{knigge06} Knigge C. 2006, MNRAS, 373, 484

\bibitem[Kubota et al.(1998)]{kubota98} Kubota A., Tanaka Y., Makishima K., Ueda Y., Dotani T., Inoue H., Yamaoka K., 1998, PASJ, 50, 667

\bibitem[Kuulkers et al.(2013)]{kuulkers13} Kuulkers E., et al., 2013, A\&A, 552, 32

\bibitem[\protect\citeauthoryear{Lasota}{2001}]{lasota01} Lasota J.-P., 2001, NewAR, 45, 449%. doi:10.1016/S1387-6473(01)00112-9

\bibitem[\protect\citeauthoryear{Lasota et al.}{2015}]{lasota15} Lasota J.-P., King A.R., Dubus G., 2015, ApJ, 801, L4

\bibitem[Lazar et al.(2021)]{lazar21} Lazar H., et al., 2021, ApJ, 921, 155

\bibitem[Liu et al.(2020)]{liu20} Liu J.F., et al., 2020, ApJ, 900, 42 

\bibitem[Ma et al.(2022)]{ma22} Ma R.C., et al., 2022, MNRAS, in press 
    (arXiv:2206.03409)

\bibitem[Maccarone \& Patruno(2013)]{maccarone13} Maccarone T.J., Patruno A., 2013, MNRAS, 428, 1335

\bibitem[Maccarone(2003)]{maccarone03} Maccarone T.J., 2003, A\&A, 409, 697

\bibitem[McClintock et al.(2014)]{mcclintock14} McClintock J.E., Narayan R., Steiner J.F., 2014, SSRv, 183, 295

\bibitem[Mata Sanchez et al.(2015)]{mata15} Mata Sanchez D., Mu\~noz-Darias T., Casares J., Corral-Santana J.M., Shahbaz, T., 2015, MNRAS, 454, 2199

\bibitem[Matthews et al.(2015)]{matthews15} Matthews J.H., Knigge C., Long K.S., Sim S.A., Higginbottom N., 
2015, MNRAS, 450, 3331

\bibitem[Mu\~noz-Darias et al.(2011)]{munoz-darias11}  Mu\~noz-Darias T., Motta S., Stiele H., Belloni T.M., 2011, MNRAS, 415, 292

\bibitem[Murray \& Chiang(1996)]{murray96}  Murray N., Chiang J., 1996,  Natur, 382, 789 

\bibitem[Murray \& Chiang(1997)]{murray97}  Murray N., Chiang J., 1997, ApJ, 474, 91 

\bibitem[Negoro et al.(2019)]{negoro19} Negoro H., et al., 2019, The Astronomer’s Telegram, 13256, 1

\bibitem[Orosz et al.(1994)]{orosz94} Orosz J.A., Bailyn C.D., Remillard R.A., McClintock J.E., Foltz C.B., 1994, ApJ, 436, 848

\bibitem[Paczynski(1977)]{paczynski77} Paczynski B., 1977, ApJ, 216, 822 

\bibitem[Parsons et al.(2018)]{parsons18} Parsons S.G., et al., 2018, MNRAS, 481, 1083

\bibitem[Patterson(1984)]{patterson84} Patterson J. 1984, ApJS, 54, 443


\bibitem[Patterson et al.(2005)]{patterson05} Patterson J., et al., 2005, PASP, 117, 1204

\bibitem[Pecaut \& Mamajek(2013)]{pecaut13} Pecaut M.J., Mamajek E.E.,  
2013, ApJS, 208, 9

\bibitem[Pfahl et al.(2003)]{pfahl03} Pfahl E., Rappaport S.A.,  Podsiadlowski P., 2003, ApJ, 597, 1036



\bibitem[Podsiadlowski et al.(2002)]{podsiadlowski02}      Podsiadlowski P., Rappaport S., Pfahl E.D. 2002, ApJ, 565, 1107

\bibitem[Ponti et al.(2018)]{ponti18} Ponti G., Bianchi S., Mu\~noz-Darias T., Nandra K., 2018, MNRAS, 481, L94

\bibitem[Powell et al.(2007)]{powell07} Powell C.R., Haswell C.A., Falanga M., 
2007, MNRAS, 374, 466

\bibitem[Pylyser et al.(1988)]{pylyser88} Pylyser E., Savonije G.J., 1988, A\&A, 191, 57

\bibitem[Russell et al.(2019)]{russell19} Russell T.D., Miller-Jones J.C.A., Sivakoff G.R., Tetarenko A.J., 2019, The Astronomer’s Telegram, 13275, 1

\bibitem[Rappaport et al.(1983)]{rappaport83}  Rappaport S., Verbunt F., Joss P.C. 1983, ApJ, 275, 713 

\bibitem[Remillard et al.(2020)]{remillard20} Remillard R., Pasham D., Gendreau K., Arzoumanian Z., Homan J., Altamirano D., Steiner J., 2020, The Astronomer’s Telegram, 13427, 1

\bibitem[Repetto et al.(2017)]{repetto17}  Repetto S., Igoshev A.P., Nelemans G. 2017, MNRAS, 467, 298

\bibitem[Sakhibullin et al.(1998)]{sakhibullin98} Sakhibullin N.A., Suleimanov V.F., Shimanskii V.V., Suleimanova S.L., 1998, AstL, 24, 22

\bibitem[Shafee et al.(2006)]{shafee06} Shafee R., McClintock J.E., Narayan R., Davis S.W., Li L.-X., Remillard R.A., 2006, ApJ, 636, L113

\bibitem[Shakura \& Sunyaev(1973)]{ss73} Shakura N.I., Sunyaev R.A., 1973, A\&A, 24, 337 

\bibitem[Shimura \& Takahara(1995)]{shimura95} Shimura T., Takahara F., 1995, ApJ, 445, 780 

\bibitem[Smak(1981)]{smak81} Smak J.,  1981, AcA, 31, 395

\bibitem[Smak(1991)]{smak91} Smak J., 1991, The 6th Institute d’Astrophysique de Paris (IAP) Meeting: Structure and Emission Properties of Accretion Disks,  ed. C. Bertout, S. Collin-Souffrin, \&  J.-P. Lasota (Gif-sur-Yvette: Editions Frontieres), IAUColl., 129, 247

\bibitem[Spruit \& Ritter(1983)]{spruit83}  Spruit H.C., Ritter H.,  1983, A\&A, 124, 267

\bibitem[Strader et al.(2019)]{strader19} Strader J., Aydi E., Sokolovsky K., Shishkovsky L., 2019, The Astronomer’s Telegram, 13260, 1

\bibitem[Tetarenko et al.(2016)]{tetarenko16} Tetarenko B.E., Sivakoff G.R., Heinke C.O., Gladstone J.C., 2016, ApJS, 222, 15

\bibitem[Tetarenko et al.(2018)]{tetarenko18} Tetarenko B.E., Dubus G., Lasota J.-P., Heinke C.O., Sivakoff G.R., 2018, MNRAS, 480, 2 

\bibitem[Tetarenko et al.(2021)]{tetarenko21} Tetarenko B.E., Shaw A.W., Manrow E.R., Charles P.A., Miller J.M., Russell T.D., Tetarenko A.J.,  2021, MNRAS, 501, 3406 

\bibitem[Thomas et al.(2019)]{thomas19} Thomas N.T., Gudennavar S.B., Misra R., Bubbly S.G., 2019, ATel, 13296, 1

\bibitem[Tomsick et al.(2019)]{tomsick19} Tomsick J.A., et al., 2019, The Astronomer’s Telegram, 13270, 1

\bibitem[Torres et al.(2021)]{torres21}  Torres M.A.P., Jonker P.G., Casares J., Miller-Jones J.C.A., Steeghs D., 2021, MNRAS, 501, 2174

\bibitem[van Paradijs \& McClintock(1994)]{van94} van Paradijs J., McClintock J.E., 1994, A\&A, 290, 133

\bibitem[Warner(1995)]{warner95} Warner B., 1995, Cataclysmic Variable Stars, (Cambridge UK: Cambridge Univ. Press)

\bibitem[Watson(2011)]{watson11} Watson D., 2011, A\&A 533, A16

\bibitem[Webbink \& Wickramasinghe(2002)]{webbink02}   Webbink R.F., Wickramasinghe D.T.  2002, MNRAS, 335, 1

\bibitem[Whitehurst \& King(1991)]{whitehurst91} Whitehurst R., King A., 1991, MNRAS, 249, 25

\bibitem[Wickramasinghe \& Hubeny(2005)]{wickramasinghe05}      Wickramasinghe D.T., Hubeny I., 2005, ASPC, 330, 219 

\bibitem[Willingale et al.(2013)]{willingale13} Willingale R., Starling R.L.C., Beardmore A.P., Tanvir N.R., O'Brien P.T., 2013, MNRAS, 431, 394

\bibitem[Wu et al.(2001)]{wu01}  Wu K., Soria R., Hunstead R.W., Johnston H.M., 2001, MNRAS, 320, 177

\bibitem[Yamaoka et al.(2012)]{yamaoka12} Yamaoka K., et al., 2012, PASJ, 64, 32

\bibitem[Zhang et al.(1997)]{zhang97}  Zhang S.-N., Cui W., Chen W., 1997, ApJ, 482, L155

\bibitem[Zhang et al.(2012)]{zhang12}  Zhang S.-N., Liao J.,  Yao Y., 
2012, MNRAS, 421, 3550 

\end{thebibliography}
\end{document}